\pdfoutput=1
\documentclass[]{osa-article}

\journal{osajournal}

\articletype{Research Article}

\usepackage{empheq}
\usepackage{bm}
\usepackage[normalem]{ulem}
\usepackage{booktabs}

\let\originalleft\left
\let\originalright\right
\renewcommand{\left}{\mathopen{}\mathclose\bgroup\originalleft}
\renewcommand{\right}{\aftergroup\egroup\originalright}
\DeclareMathAlphabet{\mathcal}{OMS}{cmsy}{m}{n}

\providecommand{\mb}[1]{\mathbf{#1}}
\providecommand{\msf}[1]{\mathsf{#1}}

\providecommand{\mc}[1]{\mathcal{#1}}
\providecommand{\ro}{\mathbf{\mathbf{r}}_o}
\providecommand{\so}{\mathbf{\hat{s}}_o}
\providecommand{\rp}{\mathbf{r}_p}
\providecommand{\rbm}[1]{r_b^{\text{m}}}
\providecommand{\rd}{\mathbf{r}_d}

\providecommand{\mh}[1]{\mathbf{\hat{#1}}}
\providecommand{\mbb}[1]{\mathbb{#1}}
\providecommand{\bs}[1]{\boldsymbol{#1}}
\providecommand{\bv}{\bs{\nu}}
\providecommand{\taup}{\bs{\tau}}

\begin{document}

\title{Spatio-angular fluorescence microscopy\\ II. Paraxial 4f imaging}

\author{Talon Chandler,\authormark{1,*} Hari Shroff,\authormark{2,3} Rudolf Oldenbourg,\authormark{3} and Patrick La Rivi\`ere\authormark{1,3}}

\address{\authormark{1}University of Chicago, Department of Radiology, Chicago, Illinois 60637, USA\\
  \authormark{2}Section on High Resolution Optical Imaging, National Institute
  of Biomedical Imaging and Bioengineering, National Institutes of Health,
  Bethesda, Maryland 20892, USA\\
  \authormark{3}Marine Biological Laboratory, Bell Center, Woods Hole, Massachusetts 02543, USA
}

\email{\authormark{*}talonchandler@talonchandler.com} 



\begin{abstract*}
  We investigate the properties of a single-view fluorescence microscope
  in a $4f$ geometry when imaging fluorescent dipoles without using the
  monopole or scalar approximations. We show that this imaging system has a
  spatio-angular band limit, and we exploit the band limit to perform efficient
  simulations. Notably, we show that information about the out-of-plane
  orientation of ensembles of in-focus fluorophores is recorded by paraxial
  fluorescence microscopes. Additionally, we show that the monopole
    approximation may cause biased estimates of fluorophore concentrations, but
  these biases are small when the sample contains either many randomly oriented
  fluorophores in each resolvable volume or unconstrained rotating fluorophores.
\end{abstract*}

\section{Introduction}
In the first paper of this series we developed a new set of transfer
  functions that can be used to analyze spatio-angular fluorescence microscopes
\cite{chandler2018}. In this work we will demonstrate these transfer
  functions by analyzing a single-view fluorescence microscope in a $4f$
  geometry.

A central goal of this work is to examine the validity of the monopole
  approximation in fluorescence microscopy. Although many works implicitly
apply the monopole approximation, we have encountered two explicit
justifications: (1) the sample contains many randomly oriented fluorophores
within a resolvable volume or (2) the sample contains unconstrained rotating
fluorophores. While both of these situations yield monopole-like emitters,
neither yields emitters that are perfectly described by the monopole model. We
investigate the dipole model of fluorophores in detail and find the conditions
under which the monopole approximation is justified.

We begin in section \ref{sec:theory} by specifying the imaging geometry and
defining \textit{pupil functions} for imaging systems with and without the
monopole approximation. We explicitly relate the pupil functions to the coherent
transfer functions to establish a connection between physical calculations and
the transfer functions. Next, in section \ref{sec:results} we calculate the
monopole and dipole transfer functions in closed form, and we use these transfer
functions to perform efficient simulations with four numerical phantoms.
Finally, in section \ref{sec:discussion} we discuss the results and expand on
how the pupil functions can be used to develop improved models for
spatio-angular microscopes.

\section{Theory}\label{sec:theory}
During our initial modeling \cite{chandler2018} we considered an
  aplanatic optical system imaging a sample of in-focus fluorophores---either a
  monopole density, $f(\ro)$, or a dipole density, $f(\ro, \so)$---by recording
  the scaled irradiance on a two-dimensional detector, $g(\rd)$. A central
  result was that we could express the relationship between the object and the
  data as a linear Hilbert-space operator, and we showed that these operators
  took the form of an integral transform in a delta function basis. For
  monopoles the integral transform takes the form
\begin{align}
  g(\rd) = \int_{\mbb{R}^2}d\ro\,h(\rd - \ro)f(\ro), \label{eq:mono}
\end{align}
where $h(\rd - \ro)$ is the monopole point spread function. For dipoles the
  integral transform takes the form
\begin{align}
  g(\rd) = \int_{\mbb{S}^2}d\so\int_{\mbb{R}^2}d\ro\,h(\rd - \ro,\so)f(\ro,\so),\label{eq:dip}
\end{align}
where $h(\rd - \ro, \so)$ is the dipole point spread function. Note that we have
written Eqs. \eqref{eq:mono} and \eqref{eq:dip} in their demagnified forms. We
will use primes to denote the unscaled detector coordinate, $\rd'$, and unscaled
point spread functions, $h'$.

After expressing the operators in a delta function basis we explored the
  form of the operators with several other choices of basis functions. Tables 1
  and 2 summarize our results.

  \begin{table}[t]
    \centering
  \begin{tabular}{lll}
    \toprule
    Quantity&Symbol&Relationships\\
    \midrule
    Monopole density&$f(\ro)$&---\\
    Monopole spectrum&$F(\bv)$&$=\mc{F}_{\mbb{R}^2}\left\{f(\ro)\right\}$\\
    \midrule
    Monopole coherent spread function&$c(\rd - \ro)$&---\\
    Monopole coherent transfer function&$C(\taup)$&$=\mc{F}_{\mbb{R}^2}\left\{c(\rd - \ro)\right\}$\\
    Monopole point spread function&$h(\rd - \ro)$&$=|c(\rd - \ro)|^2$\\
    Monopole transfer function&$H(\bv)$&$=\mc{F}_{\mbb{R}^2}\left\{h(\rd-\ro)\right\}$\\
    &&$= \int_{\mbb{R}^2}d\taup\,C(\taup)C^{*}(\taup - \bv)$\\
    \midrule
    Scaled irradiance&$g(\rd)$&$=\int_{\mbb{R}^2}d\ro\, h(\rd - \ro)f(\ro)$\\
    Scaled irradiance spectrum&$G(\bv)$&$=\mc{F}_{\mbb{R}^2}\left\{g(\rd)\right\}=H(\bv)F(\bv)$\\
    \bottomrule
\end{tabular}
\caption{Summary of relevant quantities in fluorescence microscopy under the
  monopole approximation---see \cite{chandler2018} for derivations.
  $\mc{F}_{\mbb{R}^2}$ denotes a two-dimensional Fourier transform.}
\end{table}

  \begin{table}[t]
    \hspace{-1.5em}
  \begin{tabular}{lll}
    \toprule
    Quantity&Symbol&Relationships\\
    \midrule
    Dipole density&$f(\ro,\so)$&---\\
    Dipole spatial spectrum&$F(\bv,\so)$&$=\mc{F}_{\mbb{R}^2}\left\{f(\ro,\so)\right\}$\\
    Dipole angular spectrum&$F_{\ell}^m(\ro)$&$=\mc{F}_{\mbb{S}^2}\left\{f(\ro,\so)\right\}$\\
    Dipole spatio-angular spectrum&$\msf{F}_{\ell}^m(\bv)$&$=\mc{F}_{\mbb{R}^2}\left\{F_{\ell}^m(\ro)\right\} = \mc{F}_{\mbb{S}^2}\left\{F(\bv, \so)\right\}$\\
    \midrule
    Dipole coherent spread function&$\mb{c}(\rd - \ro, \so)$&---\\
    Dipole coherent transfer function&$\mb{C}(\taup,\so)$&$=\mc{F}_{\mbb{R}^2}\left\{\mb{c}(\rd - \ro, \so)\right\}$\\
    Dipole point spread function&$h(\rd - \ro,\so)$&$=|\mb{c}(\rd - \ro, \so)|^2$\\
    Dipole spatial transfer function&$H(\bv,\so)$&$=\mc{F}_{\mbb{R}^2}\left\{h(\rd-\ro,\so)\right\}$\\
    &&$= \int_{\mbb{R}^2}d\taup\,\mb{C}(\taup,\so)\mb{C}^{\dagger}(\taup - \bv,\so)$\\
    Dipole angular transfer function&$H_{\ell}^m(\rd - \ro)$&$=\mc{F}_{\mbb{S}^2}\left\{h(\rd-\ro,\so)\right\}$\\
    Dipole spatio-angular transfer function&$\msf{H}_{\ell}^m(\bv)$&$=\mc{F}_{\mbb{R}^2}\left\{H_{\ell}^m(\rd - \ro)\right\} = \mc{F}_{\mbb{S}^2}\left\{H(\bv, \so)\right\}$\\
    \midrule
    Scaled irradiance&$g(\rd)$&$=\int_{\mbb{S}^2}d\so\int_{\mbb{R}^2}d\ro\, h(\rd - \ro, \so)f(\ro,\so)$\\
    &&$= \sum_{\ell m}\int_{\mbb{R}^2}d\ro\,H_{\ell}^m(\rd - \ro)F_{\ell}^m(\ro)$\\
    Scaled irradiance spectrum&$G(\bv)$&$=\mc{F}_{\mbb{R}^2}\left\{g(\rd)\right\}$\\
            &&$=\int_{\mbb{S}^2}d\so\, H(\bv,\so)F(\bv, \so)$\\
            &&$=\sum_{\ell m} \msf{H}_{\ell}^m(\bv) \msf{F}_{\ell}^m(\bv)$\\
    \bottomrule
\end{tabular}
  \caption{Summary of relevant quantities in spatio-angular dipole imaging---see \cite{chandler2018} for derivations. $\mc{F}_{\mbb{R}^2}$ denotes a two-dimensional Fourier transform, and $\mc{F}_{\mbb{S}^2}$ denotes a spherical Fourier transform.}
\end{table}

\begin{figure}[ht]
 \centering
   \centering
   \includegraphics[width = 1.0\textwidth]{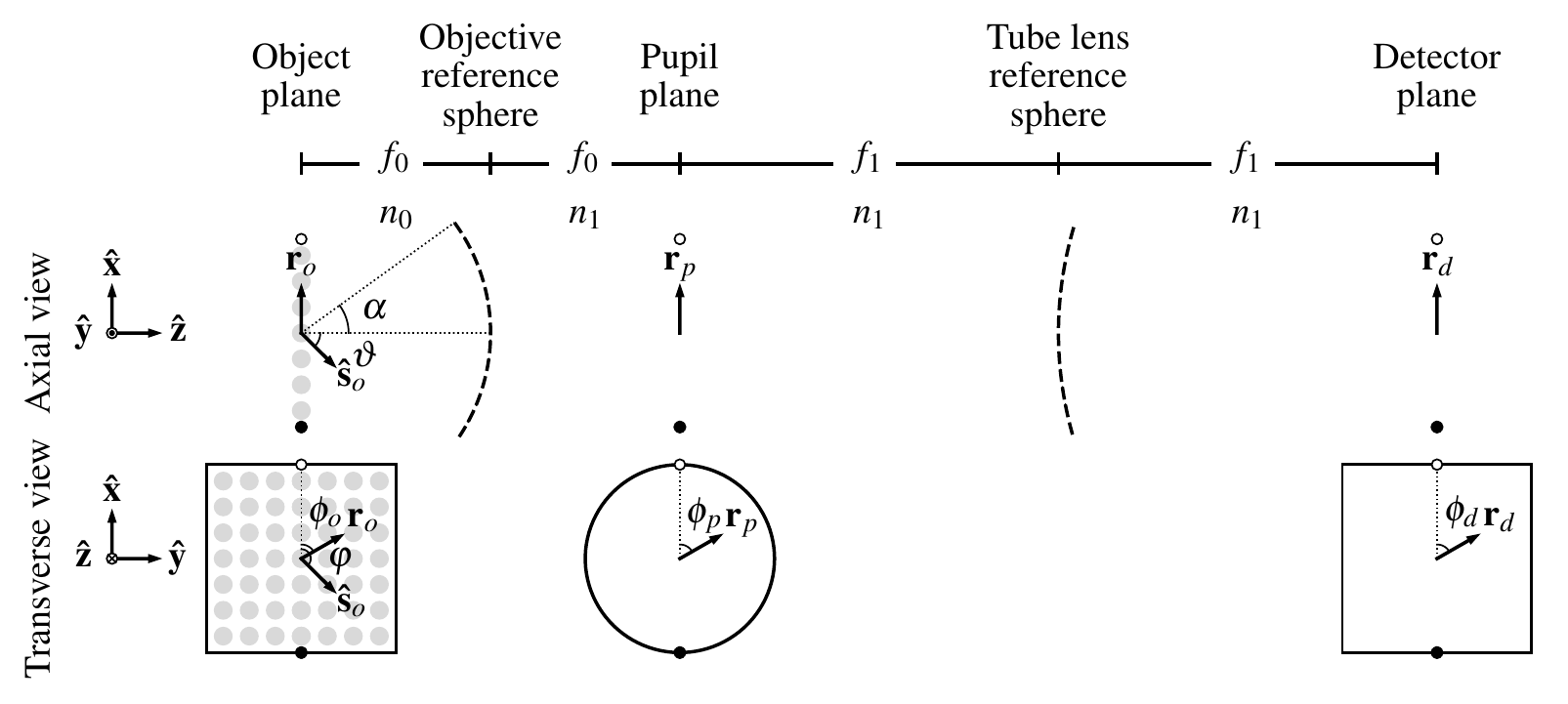}
   \caption{Schematic of an aplanatic imaging system in a $4f$ geometry with a
     paraxial tube lens. We are considering an aplanatic optical system, so we
     only need to consider the image created by on-axis objects. The fluorescent
     object consists of ensembles of monopoles or dipoles embedded in a
     medium with index of refraction $n_0$. An objective with focal length $f_0$
     and numerical aperture $\text{NA} = n_o\sin\alpha$ is trained on the
     object. A paraxial tube lens with focal length $f_1$ and a detector
     complete the $4f$ geometry, and all components except the object are
     embedded in a medium with index of refraction $n_1$. The object, pupil, and
     detector planes are parameterized by vectors $\ro$, $\rp$, and $\rd$ with
     polar coordinates ($r_o, \phi_o$), ($r_b, \phi_b$), and ($r_d, \phi_d$),
     respectively. At each position $\ro$ in the object there is a sphere
     parameterized by a unit vector $\so$ with spherical coordinates
     ($\vartheta, \varphi$). }
   \label{fig:schematic}
 \end{figure}

Our task is to calculate the form of the monopole and dipole transfer
  functions for a specific imaging geometry. In this work we will consider an
aplanatic optical system in a \textit{$\mathit{4}f$ configuration} with an
arbitrary first lens (the objective lens) and a \textit{paraxial second lens}
(the tube lens) as shown in Fig. \ref{fig:schematic}. A lens can be considered
paraxial if the angle $\alpha$ between the optical axis of the lens and the
marginal ray is small enough that $\sin\alpha \approx \alpha$. As a rule of
thumb, non-paraxial effects only become significant when the numerical aperture
of a lens exceeds 0.7 \cite[ch.~6]{gu2000}, but this is only a rough guideline.
Commercial microscopes with infinity-corrected objectives can almost always can
be modeled by considering the tube lens as paraxial.

\subsection{Monopole pupil functions}
We define the \textit{monopole pupil function} $p(\rp)$ of the imaging system as
the field immediately following the pupil plane created by an on-axis monopole,
where $\rp$ is an unscaled two-dimensional coordinate in the pupil plane. In
this section we will relate the monopole pupil function to the monopole transfer
functions by adapting the treatment in Barrett and Myers
\cite[ch.~9.7]{barrett2004}.

Since monopoles emit scalar fields, the monopole pupil function is a
scalar-valued function. The optical system is aplanatic, so we can write the
field, $U_p(\rp, \ro)$, created at a point in the pupil plane $\rp$ by a
monopole at position $\ro$ as
\begin{align}
   U_p(\rp, \ro) \propto p(\rp)\,\text{exp}\left[-2\pi i \frac{n_0}{\lambda f_0} \rp\cdot\ro \right], \label{eq:pupil}
\end{align}
where $\lambda$ is the emission wavelength. Equation \eqref{eq:pupil} is a
restatement of the aplanatic condition for a $4f$ optical system---the fields in
the pupil plane can be written as the pupil function multiplied by a linear
phase factor that encodes the position of the object.

Since the second lens is paraxial, we can model the relationship between the
field in the pupil plane and the field on the detector with a scaled Fourier
transform \cite{goodman1996, axelrod2012, backer2014}:
\begin{align}
   U_d(\rd', \ro) \propto \int_{\mbb{R}^2}d\rp\, p(\rp)\,\text{exp}\left[-2\pi i \frac{n_0}{\lambda f_0} \rp\cdot\ro \right]\text{exp}\left[-2\pi i \frac{n_1}{\lambda f_1} \rp\cdot\rd' \right], \label{eq:detectorfourier}
\end{align}
where $\rd'$ is an unscaled detector coordinate.

If we define $P(\taup)$ as the two-dimensional Fourier transform of the pupil function
then we can rewrite Eq. \eqref{eq:detectorfourier} as
\begin{align}
  U_d(\rd', \ro) \propto P\left(\frac{n_0}{\lambda f_0}\ro + \frac{n_1}{\lambda f_1}\rd'\right),
\end{align}
which we can simplify further by writing in terms of the magnification
$m = -\frac{f_1n_0}{f_0n_1}$:
\begin{align}
  U_d(\rd' - m\ro) \propto P\left(\frac{n_1}{\lambda f_1}[\rd' - m\ro]\right).
\end{align}
The irradiance on the detector is the absolute square of the field so
\begin{align}
  h'(\rd' - m\ro) \propto \left|P\left(\frac{n_1}{\lambda f_1}[\rd' - m\ro]\right)\right|^2.
\end{align}
If we demagnify the coordinates with $\rd = \rd'/m$ and demagnify the irradiance
with $h(\rd - \ro) \propto h'(m[\rd - \ro])$, we find that the monopole point
spread function is related to the Fourier transform of the monopole pupil
function by
\begin{align}
  h(\rd - \ro) \propto \left|P\left(-\frac{n_o}{\lambda f_o}[\rd - \ro]\right)\right|^2.
\end{align}
The monopole point spread function is the absolute square of the monopole
coherent spread function so
\begin{align}
  c(\rd - \ro) \propto P\left(-\frac{n_o}{\lambda f_o}[\rd - \ro]\right).
\end{align}
Finally, the monopole coherent transfer function is the Fourier transform of the
monopole coherent spread function so
\begin{align}
  C(\taup) \propto p\left(\frac{\lambda f_o}{n_o}\taup\right). \label{eq:monopupil0}
\end{align}
Equation \eqref{eq:monopupil0} is the key result of this section---the monopole
coherent transfer function is a scaled monopole pupil function.

\subsection{Dipole pupil function}
We define the \textit{dipole pupil function} $\mb{p}(\rp, \so)$ of the imaging
system as the electric field immediately following the pupil plane created by
an on-axis dipole oriented along $\so$. Since dipoles emit vector-valued
electric fields, the dipole pupil function is a vector-valued function. Almost
all of the arguments in the previous section carry over to the dipole case.
Briefly, we can write the electric field created at a point in the pupil
$\rp$ by a dipole at $\ro$ oriented along $\so$ as
 \begin{align}
   \mb{E}_p(\rp, \ro, \so) \propto \mb{p}(\rp, \so)\,\text{exp}\left[-2\pi i \frac{n_0}{\lambda f_0} \rp\cdot\ro \right]. \label{eq:pupil2}
 \end{align}
 The second lens is paraxial, so we can find the field on the detector with a
 Fourier transform
 \begin{align}
   \mb{E}_d(\rd', \ro, \so) \propto \int_{\mbb{R}^2}d\rp\, \mb{p}(\rp, \so)\,\text{exp}\left[-2\pi i \frac{n_0}{\lambda f_0} \rp\cdot\ro \right]\text{exp}\left[-2\pi i \frac{n_1}{\lambda f_1} \rp\cdot\rd' \right]. \label{eq:detectorfourier2}
\end{align}
Note that the Fourier transform of a vector field is the Fourier transform of
its scalar-valued orthogonal components, so Eq. \eqref{eq:detectorfourier2}
specifies three two-dimensional Fourier transforms. We follow the same
manipulations as the previous section and find that the dipole coherent transfer
function is a scaled dipole pupil function
\begin{align}
  \mb{C}(\taup, \so) \propto \mb{p}\left(\frac{\lambda f_o}{n_o}\taup, \so\right). \label{eq:monopupildip}
\end{align}

We have restricted our analysis to paraxial tube lenses, but non-paraxial tube
lenses (or a non-infinity-corrected objective) can be modeled with vector-valued
three-dimensional pupil functions \cite{sheppard1994, gu2000, arnison2002,
  foreman2011-2}.

\subsection{Special functions}
We adopt and generalize Bracewell's notation \cite{bracewell2004} for several
special functions which will simplify our calculations. First, we define a
\textit{rectangle function} as
\begin{align}
  \Pi(x) = 
  \begin{cases}
    1\quad \text{if}\quad |x| < \frac{1}{2},\\
    0\quad \text{else}.
  \end{cases}
\end{align}
We also define the \textit{$n^{th}$-order jinc function} as
\begin{align}
  \text{jinc}_n(r) = \frac{J_{n+1}(\pi r)}{2r},
\end{align}
where $J_{n+1}(r)$ is the $(n+1)^{th}$-order Bessel function of the first kind.

Although the rectangle and jinc functions are defined in one dimension, we will
usually apply them in two dimensions. In Appendix \ref{sec:special} we derive
the following two-dimensional Fourier transform relationships between the jinc
functions and the weighted rectangle functions
\begin{align}
  i^{n}\left\{\substack{
  \text{exp}(in\phi_r)\\
    \cos(n\phi_r)\\
    \sin(n\phi_r)
  }\right\}\text{jinc}_n(r) &\stackrel{\mathcal{F}_{\mbb{R}^2}}{\longrightarrow} (2\nu)^{n}\left\{\substack{
    \text{exp}(in\phi_{\nu})\\
    \cos(n\phi_{\nu})\\
    \sin(n\phi_{\nu})
  }\right\}\Pi(\nu), \label{eq:jincrect2}
  \end{align}
  where the entries inside the curly braces are to be taken one at a time and
  $\{r, \phi_r\}/\{\nu, \phi_\nu\}$ are conjugate sets of polar coordinates.
  
Finally, we define the \textit{$n^{th}$-order chat function} as the
two-dimensional Fourier transform of the squared $n^{th}$-order jinc function
\begin{align}
  \text{jinc}_n^2(r) \stackrel{\mathcal{F}_{\mbb{R}^2}}{\longrightarrow} \text{chat}_n(\nu).
\end{align}

In Appendix \ref{sec:special} we show that the zeroth- and first-order chat
functions can be written in closed form as
\begin{align}
  \text{chat}_0(x) &=  \frac{1}{2}\left[\cos^{-1}|x| - |x|\sqrt{1 - x^2}\right]\Pi\left(\frac{x}{2}\right),\\
  \text{chat}_1(x) &=  \frac{1}{2}\left[\cos^{-1}|x| - |x|(3-2x^2)\sqrt{1 - x^2}\right]\Pi\left(\frac{x}{2}\right).  
\end{align}

\section{Results}\label{sec:results}
\subsection{Monopole transfer functions}
Our first step towards the monopole transfer functions is to calculate the
monopole pupil function and coherent transfer function. Several works
\cite{petrov2017, backlund2018} have modeled an aplanatic fluorescence
microscope imaging monopole emitters with the scalar pupil function
\begin{align}
  p(\rp) \propto \tilde{C}\left(\frac{r_p}{f_o}\right)\Pi\left(\frac{r_p}{2f_o\sin\alpha}\right), 
\end{align}
where
\begin{align}
  \tilde{C}(x) = (1 - x^2)^{-1/4} = 1 + \frac{x^2}{4} + \frac{5x^4}{32} + \cdots. 
\end{align}
The $\tilde{C}(x)$ function models the radial dependence of the field and
ensures that power is conserved on either side of an aplanatic objective, and
the rectangle function models the aperture stop of the objective. Applying Eq.
\eqref{eq:monopupil0} and collecting constants we find that the coherent monopole
transfer function is
\begin{align}
  C(\taup) \propto \tilde{C}\left(\frac{2\text{NA}}{n_o}\frac{\tau}{\nu_c}\right)\Pi\left(\frac{\tau}{\nu_c}\right), \label{eq:coherentmonopole}
\end{align}
where $\text{NA} = n_o\sin\alpha$ and $\nu_c = 2\text{NA}/\lambda$. This
coherent transfer function models objectives with an arbitrary numerical
aperture, but for our initial analysis we restrict ourselves to the paraxial
regime. We drop second- and higher-order radial terms to find that
\begin{align}
  C(\taup) \stackrel{(p)}{\propto} \Pi\left(\frac{\tau}{\nu_c}\right), 
\end{align}
where $(p)$ indicates that we have used the paraxial approximation for the
objective lens.

We can find the monopole coherent spread function by taking the inverse Fourier
transform of the monopole coherent transfer function
\begin{align}
  c(\mb{r}) \stackrel{(p)}{\propto} \text{jinc}_0(\nu_c r).
\end{align}

The monopole point spread function is the (normalized) absolute square of the
monopole coherent spread function so
\begin{align}
  h(\mb{r}) \stackrel{(p)}{=} \frac{4}{\pi}\text{jinc}_0^2(\nu_c r),
\end{align}
which is the well-known \textit{Airy disk}.

Finally, we can calculate the monopole transfer function as the two-dimensional Fourier transform of the monopole point spread function (or the autocorrelation
of the coherent transfer function) and find that 
\begin{align}
  H(\bs{\nu}) \stackrel{(p)}{=} \frac{4}{\pi}\text{chat}_0\left(\frac{\nu}{\nu_c}\right).
\end{align}

 \subsection{Dipole transfer functions}
 To calculate the dipole transfer function we proceed similarly to the monopole
 case---we find the pupil function, scale to find the coherent dipole transfer
 function, then calculate the remaining transfer functions.
 
 Backer and Moerner \cite{backer2014} have calculated the dipole pupil function
 for a high-NA objective as
 \begin{align}
   \mb{p}(\rp, \so) \propto
   \begin{bmatrix}
     \tilde{C}_0\left(\frac{r_p}{f_o}\right) + \tilde{C}_2\left(\frac{r_p}{f_o}\right)c(2\phi_p)&\tilde{C}_2\left(\frac{r_p}{f_o}\right)s(2\phi_p)&\tilde{C}_1\left(\frac{r_p}{f_0}\right)c(\phi_p)\\
     \tilde{C}_2\left(\frac{r_p}{f_o}\right)s(2\phi_p)&\tilde{C}_0\left(\frac{r_p}{f_o}\right) - \tilde{C}_2\left(\frac{r_p}{f_o}\right)c(2\phi_p)&\tilde{C}_1\left(\frac{r_p}{f_0}\right)s(\phi_p)\\     
     0&0&0\\     
   \end{bmatrix}
   \begin{bmatrix}
     s_x\\
     s_y\\
     s_z
   \end{bmatrix}
   \Pi\left(\frac{r_p}{2f_os(\alpha)}\right),
 \end{align}
 where $c(x)$ and $s(x)$ are shorthand for $\cos(x)$ and $\sin(x)$,
 $\{s_x, s_y, s_z\}$ are the Cartesian components of $\so$ when $\mh{z}$ is
 aligned with the optical axis, and
 \begin{alignat}{4}
   \tilde{C}_0(x) &= \frac{1}{2}\left(\sqrt{1 - x^2} + 1\right)(1 - x^2)^{-1/4} &&= 1 + \frac{x^4}{32} + \frac{x^6}{32} + \cdots,\\
   \tilde{C}_1(x) &= x(1 - x^2)^{-1/4} &&= x + \frac{x^3}{4} + \frac{5x^5}{32} + \cdots,\\
   \tilde{C}_2(x) &= \frac{1}{2}\left(\sqrt{1 - x^2} - 1\right)(1 - x^2)^{-1/4} &&= -\frac{x^2}{4} - \frac{x^4}{8} - \frac{11x^6}{128} - \cdots.
 \end{alignat}
 Similar to the monopole case, the dipole pupil function conserves power and has
 a cutoff at the objective aperture, but the dipole pupil function is
 vector-valued to model the complete electric field in the pupil plane. The
 fields in the pupil plane have a negligible $\mh{z}$ component which is a
 consequence of our assumption that the tube lens is paraxial---modeling a
 non-paraxial tube lens would require a three-dimensional vector-valued pupil
 function \cite{sheppard1994, gu2000, arnison2002, foreman2011-2}. 

 Scaling the dipole pupil function using Eq. \eqref{eq:monopupildip} yields the
 dipole coherent transfer function
 \begin{equation}
   \mb{C}(\taup, \so)\hspace{-0.2em}\propto\hspace{-0.2em}
   \begin{bmatrix}
     \tilde{C}_0\left(\frac{\lambda r_p\tau}{n_0}\right)\!+\!\tilde{C}_2\left(\frac{\lambda r_p\tau}{n_0}\right)c(2\phi_{\tau})&\tilde{C}_2\left(\frac{\lambda r_p\tau}{n_0}\right)s(2\phi_{\tau})&\tilde{C}_1\left(\frac{\lambda r_p\tau}{n_0}\right)c(\phi_{\tau})\\
     \tilde{C}_2\left(\frac{\lambda r_p\tau}{n_0}\right)s(2\phi_{\tau})&\tilde{C}_0\left(\frac{\lambda r_p\tau}{n_0}\right)\!-\!\tilde{C}_2\left(\frac{\lambda r_p\tau}{n_0}\right)c(2\phi_{\tau})&\tilde{C}_1\left(\frac{\lambda r_p\tau}{n_0}\right)s(\phi_{\tau})\\     
     0&0&0\\     
   \end{bmatrix}
   \begin{bmatrix}
     s_x\\
     s_y\\
     s_z
   \end{bmatrix}
   \Pi\left(\frac{\tau}{\nu_c}\right). 
 \end{equation}
 We restrict our analysis to the paraxial regime by dropping second- and
 higher-order radial terms to find that
 \begin{align}
   \mb{C}(\taup, \so) \stackrel{(p)}{\propto}
   \begin{bmatrix}
     1&0&\frac{\text{2NA}}{n_o}\frac{\tau}{\nu_c}\cos\phi_{\tau}\\
     0&1&\frac{\text{2NA}}{n_o}\frac{\tau}{\nu_c}\sin\phi_{\tau}\\
     0&0&0\\     
   \end{bmatrix}
   \begin{bmatrix}
     s_x\\
     s_y\\
     s_z
   \end{bmatrix}
   \Pi\left(\frac{\tau}{\nu_c}\right). \label{eq:coherentdip}
 \end{align}
 Under the paraxial approximation the transverse components of the dipole
 \{$s_x, s_y$\} create purely transverse fields in the pupil plane and the axial
 component of the dipole \{$s_z$\} creates purely radial fields in the pupil
 plane. The paraxial approximation may seem crude compared to Backer and
 Moerner's numerical results, but the approximation will allow us to calculate
 the transfer functions in closed form so that we can build an intuition for the
 limits of the microscope. We also note that many existing works in ensemble
 polarized fluorescence microscopy make stronger approximations than ours. For
 example, Fourkas only considers the total irradiance in the pupil plane while
 ignoring the propagation of fields to the detector \cite{fourkas2001}.

 The dipole coherent spread function is the inverse Fourier transform of the
 dipole coherent transfer function. Applying Eq. \eqref{eq:jincrect2} in reverse
 yields
  \begin{align}
   \mb{c}(\mb{r}, \so) \stackrel{(p)}{\propto}
   \begin{bmatrix}
     \text{jinc}_0(\nu_c r)&0&\frac{\text{NA}}{n_o}i\cos\phi\,\text{jinc}_1(\nu_c r)\\
     0&\text{jinc}_0(\nu_c r)&\frac{\text{NA}}{n_o}i\sin\phi\,\text{jinc}_1(\nu_c r)\\
     0&0&0\\     
   \end{bmatrix}
   \begin{bmatrix}
     s_x\\
     s_y\\
     s_z
   \end{bmatrix}.
  \end{align}
  Notice that the radial component of the dipole coherent spread function has a
  $\pi/2$ phase shift relative to the transverse component. This phase factor
  arises because the Fourier transform of a real and odd function is purely
  imaginary.
  
  \subsubsection{Paraxial dipole point spread function}
  The dipole point spread function is the (normalized) absolute square of the
coherent dipole spread function
\begin{align}
  h(\mb{r}, \so) \propto \mb{c}(\mb{r}, \so)\mb{c}^\dagger(\mb{r}, \so). 
\end{align}
Plugging in the paraxial dipole coherent spread function and normalizing yields
\begin{align}
  h(\mb{r}, \so) \stackrel{(p)}{=} N\left[\text{jinc}_0^2(\nu_c r)\sin^2\vartheta + \left(\frac{\text{NA}}{n_o}\right)^2\text{jinc}_1^2(\nu_c r)\cos^2\vartheta\right],\label{eq:paradpsf}
\end{align}
where $\sin^2\vartheta = s_x^2 + s_y^2$, $\cos^2\vartheta = s_z^2$, and the normalization
factor is
\begin{align}
  N = 6\nu_c^2\pi^{-3/2}\left[2 + \left(\frac{\text{NA}}{n_o}\right)^2\right]^{-1}.
\end{align}

\begin{figure}[ht]
 \centering
   \centering
   \includegraphics[scale=0.8]{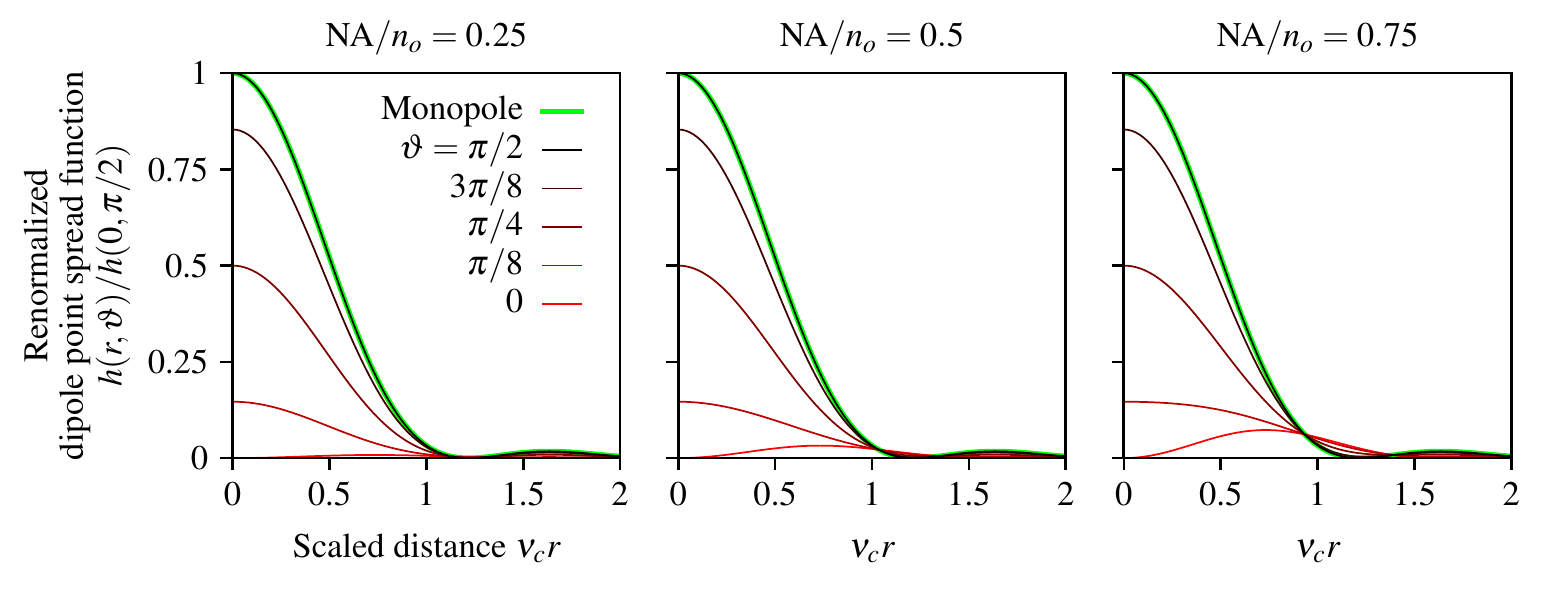}
   \caption{Renormalized paraxial dipole point spread function as a function of
     the scaled radial coordinate $\nu_c r$, the dipole inclination angle
     $\vartheta$, and $\text{NA}/n_o$. For small numerical apertures (left) the
     irradiance pattern created by axial dipoles (\textcolor{red}{\textbf{red}})
     is small compared to transverse dipoles (\textbf{black}), but the relative
     contribution of axial dipoles increases with the numerical aperture (see
     \textcolor{red}{\textbf{red}} lines from left to right). Additionally, we
     plot the monopole point spread function (\textcolor{green}{\textbf{green}})
     and observe that the paraxial monopole and dipole models are identical for
     transverse dipoles (the \textcolor{green}{\textbf{green}} and
     \textbf{black} lines are coincident).}
   \label{fig:hdet}
 \end{figure}

\begin{figure}[ht]
 \centering
   \centering
   \includegraphics[scale=0.8]{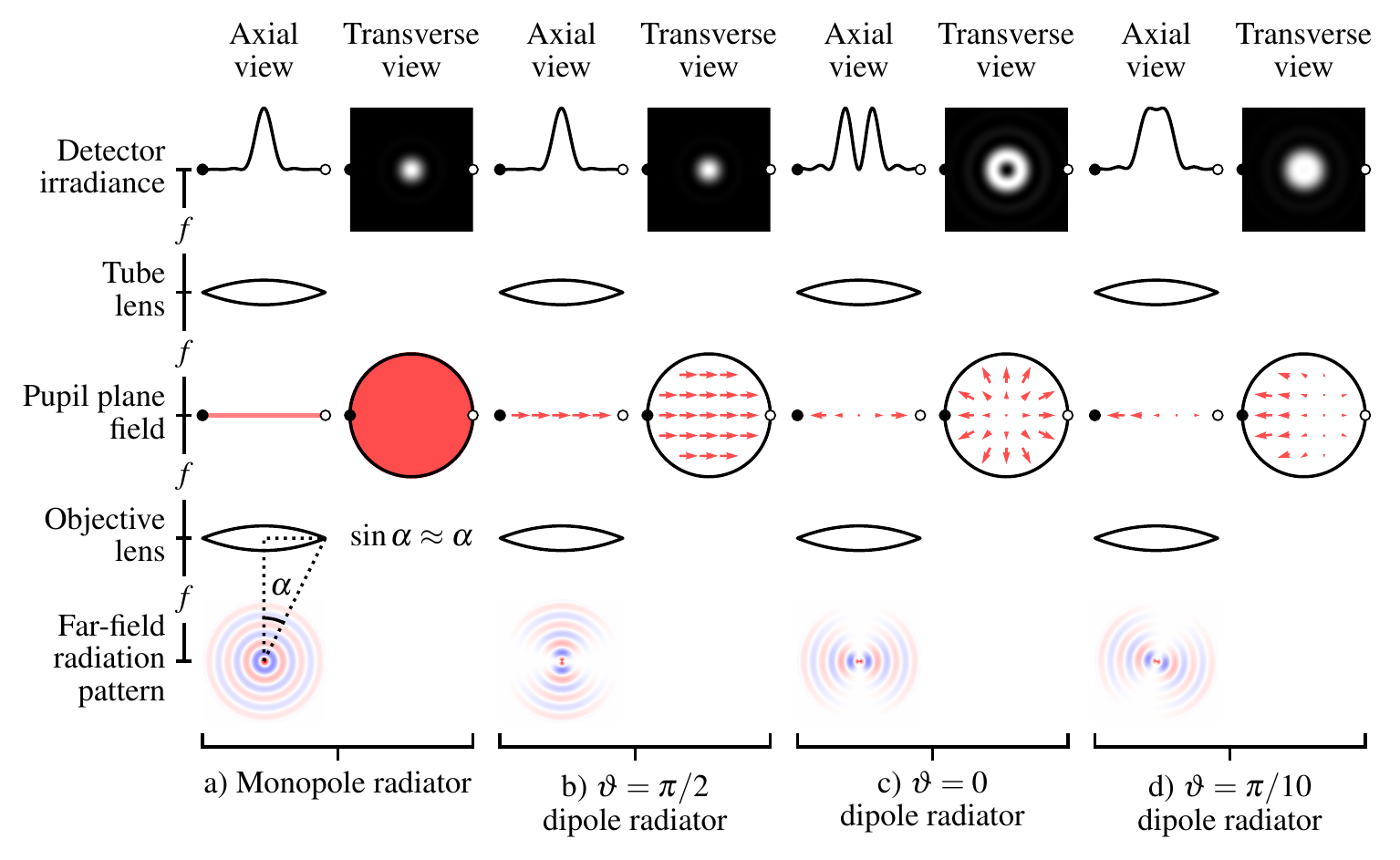}
   \caption{Comparison of paraxial models for monopole radiators a) and dipole
     radiators b)--d). a) Monopole radiators fill the pupil plane with a uniform
     scalar field which gives rise to an Airy disk on the detector. b) A
     transverse dipole radiator also creates an Airy disk, but the pupil plane
     is filled with a uniform vector field. c) An axial dipole radiator creates
     a radial electric field pattern in the back focal plane that creates a
     $\text{jinc}_1^2(r)$ pattern on the detector. d) Dipoles that are not
     transverse or axial still create radially symmetric irradiance patterns
     under the paraxial approximation. Fields from transverse dipoles are real
     and even while fields from axial dipoles are real and odd, which causes a
     relative $\pi/2$ phase shift for the fields on the detector. This phase
     shift means that the fields from transverse and axial components of the
     dipole do not interfere, which causes radially symmetric irradiance
     patterns.}
   \label{fig:microscopefig}
 \end{figure}

As discussed above, the transverse and radial fields are out of phase on the
detector, so the total irradiance is the sum of the contributions from the
transverse and radial components. In Fig. \ref{fig:hdet} we plot the dipole
point spread function for several dipole orientations and numerical apertures,
and in Fig. \ref{fig:microscopefig} we compare the monopole point spread
function to the dipole point spread function. The paraxial monopole and dipole
models are only equivalent when the sample consists of transverse dipoles, which
is clear if we notice that Eq. \eqref{eq:paradpsf} reduces to an Airy disk when
$\vartheta = \pi/2$---see Novotny and Hecht for a similar observation
\cite[ch.~4]{nov2006}.

\begin{figure}[ht]
 \centering
   \centering
   \includegraphics[scale=0.8]{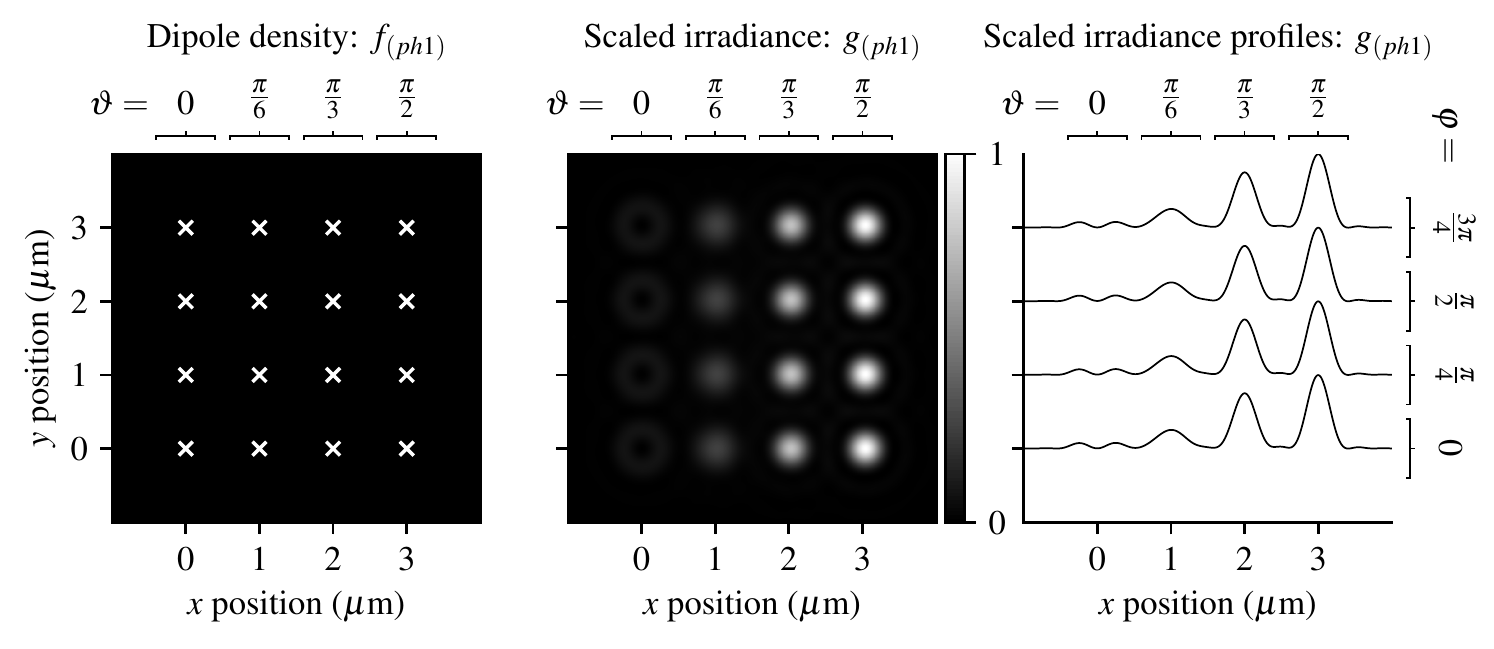}
   \caption{\textbf{Left:} A spatially and angularly sparse phantom---uniformly
     spaced single dipoles with varying orientations (increasing $\vartheta$
     from left to right and increasing $\varphi$ from bottom to top). White
     crosses mark the positions of the dipoles. \textbf{Center:} Scaled
     irradiance for an imaging system with $\text{NA} = 0.75$,
     $\lambda = 500\,\text{nm}$, and $n_o = 1.33$ sampled at $20\times$ the
     Nyquist rate. \textbf{Right:} $x$ profiles through the scaled irradiance.
     The response is independent of the azimuth angle and strongly dependent on
     the inclination angle.}
   \label{fig:ph1}
 \end{figure}

 To demonstrate the paraxial dipole point spread function we simulate a set of
 equally spaced dipoles with varying orientation:
 \begin{align}
   f_{(ph1)}(r_x, r_y, \vartheta, \varphi) = \sum_{j=0}^3 \sum_{k=0}^3 \delta\left(r_x - j\right)\, \delta\left(r_y - k\right)\, \delta\left(\cos\vartheta - \cos\vartheta_j\right)\, \delta\left(\varphi - \varphi_k\right),\label{eq:phantom1}
 \end{align}
 where $\vartheta_j = j\frac{\pi}{6}$, $\varphi_k = k\frac{\pi}{4}$, the
 subscript $(ph1)$ indicates that this is the first phantom, and the spatial
 coordinates are expressed in $\mu$m. To find the irradiance pattern created by
 the phantom we plug Eq. \eqref{eq:phantom1} into Eq. \eqref{eq:dip} and use
 the sifting property to find that
 \begin{align}
   g_{(ph1)}(r_x, r_y) = \sum_{j=0}^3 \sum_{k=0}^3 h\left(\sqrt{\left(r_x - j\right)^2 + \left(r_y - k\right)^2}, \vartheta_j\right).\label{eq:phantom1irr}
 \end{align}
 In Fig. \ref{fig:ph1} we plot the phantom and scaled irradiance for an imaging
 system with $\text{NA} = 0.75$, $\lambda = 500\,\text{nm}$, and $n_o = 1.33$.
 We sample and plot the scaled irradiance at $20\times$ the Nyquist rate,
 $\Delta x = 1/[20(2\nu_c)]$, so the irradiance patterns are free of aliasing.
 The output demonstrates that the irradiance pattern depends on the dipole
 inclination, but not its azimuth.

\subsubsection{Paraxial dipole spatial transfer function}\label{sec:trans}
\begin{figure}[ht]
 \centering
   \centering
   \includegraphics[scale=0.8]{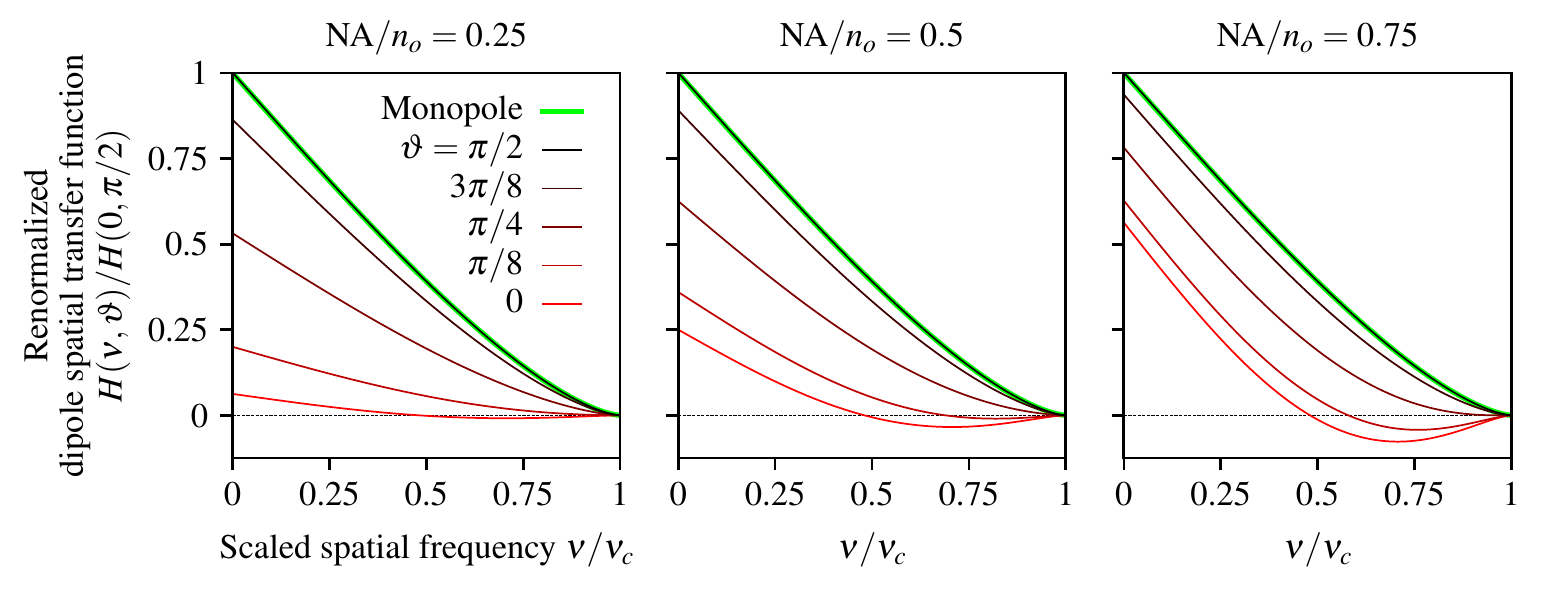}
   \caption{Dipole spatial transfer function as a function of the scaled spatial
     frequency $\nu/\nu_c$, the dipole inclination angle $\vartheta$, and
     $\text{NA}/n_o$. For small numerical apertures (left) the dipole spatial
     transfer function for axial dipoles (\textcolor{red}{\textbf{red}}) is
     small compared to transverse dipoles (\textbf{black}), but the relative
     contribution of axial dipoles increases with the numerical aperture (see
     \textcolor{red}{\textbf{red}} lines from left to right). The spatial dipole
     transfer function of axial dipoles is negative at high spatial frequencies
     because the central minimum of the axial dipole point spread function
     corresponds to the position of the dipole. Equivalently, a
     high-spatial-frequency pattern of axial dipoles will generate an irradiance
     pattern where the minimum irradiance corresponds to the peak of the axial
     dipole density. Additionally, we plot the monopole transfer function
     (\textcolor{green}{\textbf{green}}) and observe that the paraxial monopole
     and dipole models are identical for transverse dipoles (the
     \textcolor{green}{\textbf{green}} and \textbf{black} lines are
     coincident).}
   \label{fig:odotf}
 \end{figure}

The dipole spatial transfer function is the spatial Fourier transform of
the dipole point spread function (or the complex autocorrelation of the dipole
coherent transfer function). Applying the Fourier transform to Eq.
\eqref{eq:paradpsf} we find that
  \begin{align}
    H(\bs{\nu}, \vartheta) \stackrel{(p)}{=} \frac{N}{\nu_c^2}\left[\text{chat}_0\left(\frac{\nu}{\nu_c}\right)\sin^2\vartheta + \left(\frac{\text{NA}}{n_o}\right)^2 \text{chat}_1\left(\frac{\nu}{\nu_c}\right)\cos^2\vartheta\right].
  \end{align}
In Fig. \ref{fig:odotf} we plot the dipole spatial transfer function for
several dipole orientations and numerical apertures. We find that the dipole
spatial transfer function is negative for axial dipoles at high spatial
frequencies, especially for larger numerical apertures. The negative dipole
spatial transfer function corresponds to a contrast inversion for high-frequency
patterns of axial dipoles because the irradiance minimum corresponds to the
position of the dipole.

 To demonstrate the dipole spatial transfer function we simulate a set of
 equally spaced disks with varying diameter containing fluorophores with varying
 orientation
 \begin{align}
   f_{(ph2)}(r_x, r_y, \vartheta) = \sum_{j=0}^3 \sum_{k=0}^3 \frac{1}{D_k^2}\Pi\left(\frac{1}{D_k}\sqrt{\left(r_x - j\right)^2 + \left(r_y - k\right)^2}\right) \delta\left(\cos\vartheta - \cos\vartheta_j\right)\,\label{eq:phantom2}
 \end{align}
 where $D_k = 0.15(1+k)\, \mu$m and $\vartheta_j = j\frac{\pi }{6}$. Notice that
 we have scaled the disks so that the total number of fluorophores in each disk
 is constant. Also notice that the disk can model a spatial distribution of many
 fluorophores or a single molecule undergoing spatial diffusion within a well.  

 We can calculate the scaled irradiance by taking the spatial Fourier
   transform of each orientation in the phantom, multiplying the result with the
   dipole spatial transfer function, summing over the orientations, then taking
   the inverse spatial Fourier transform
 \begin{align}
   g_{(ph2)}(r_x, r_y) = \mc{F}^{-1}_{\mbb{R}^2}\left\{\sum_{j}H(\nu, \vartheta_j)\mc{F}_{\mbb{R}^2}\left\{f_{(ph2)}(r_x, r_y, \vartheta_j)\right\}\right\}.
 \end{align}

\begin{figure}[ht]
 \centering
   \centering
   \includegraphics[scale=0.8]{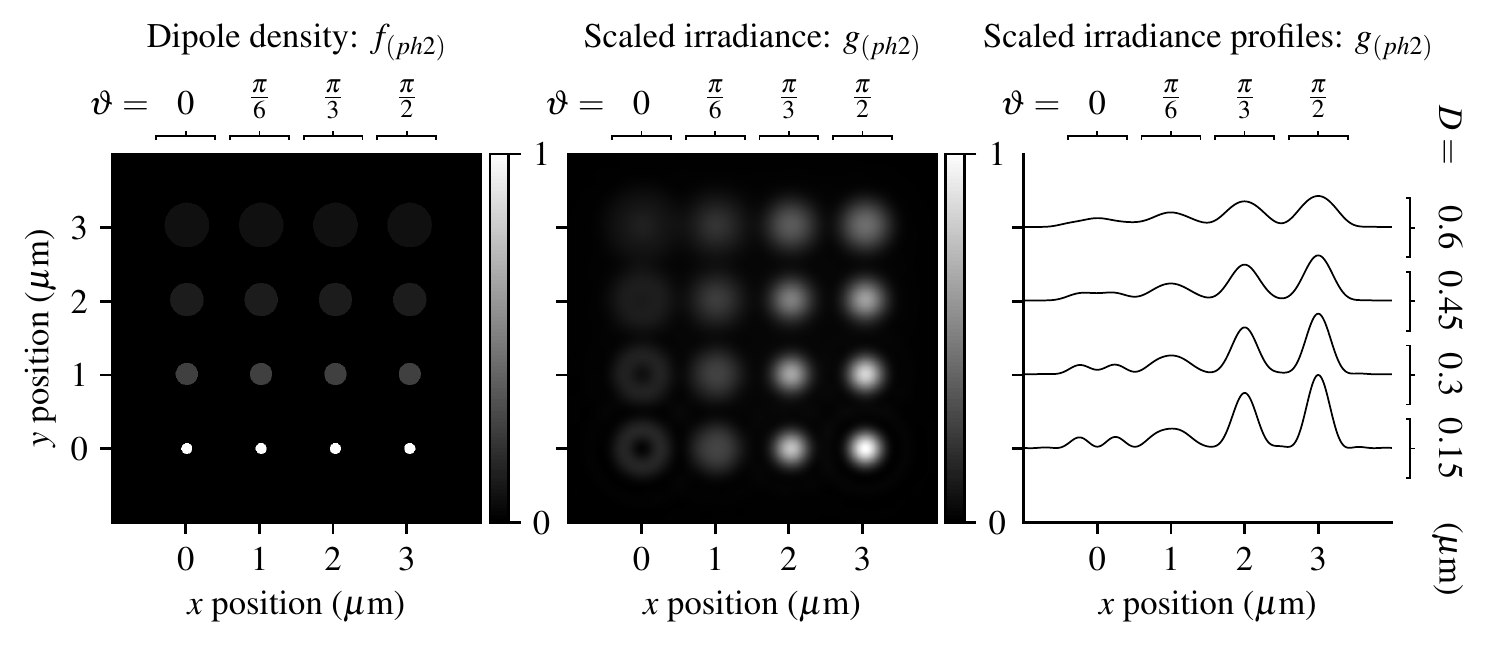}
   \caption{\textbf{Left:} A spatially dense and angularly sparse
     phantom---uniformly spaced disks with varying size (increasing $D$ from
     bottom to top) and dipole orientation (increasing $\vartheta$ from left to
     right) \textbf{Center:} Scaled irradiance for an imaging system with
     $\text{NA} = 0.75$, $\lambda = 500\,\text{nm}$, and $n_o = 1.33$ sampled at
     $20\times$ the Nyquist rate. \textbf{Right:} $x$ profiles through the
     scaled irradiance. Larger disks generate increasingly uniform irradiance
     patterns with fewer details that may indicate the orientation of
     fluorophores.}
   \label{fig:ph2}
 \end{figure}

 In Fig. \ref{fig:ph2} we plot the phantom and scaled irradiance with the same
 imaging parameters as the previous section. The small disks create irradiance
 patterns that are similar to the point sources in the previous section, while
 larger disks create increasingly uniform irradiance patterns that hide the
 orientation of the fluorophores.
  
 \subsubsection{Paraxial dipole angular transfer function}
 To calculate the angular dipole transfer function we take the spherical Fourier
 transform of the dipole point spread function
 \begin{align}
H_l^m(\mb{r}) \stackrel{}{=} \int_{\mbb{S}^2}d\so\, h(\mb{r}, \so)Y_l^{m*}(\so).
 \end{align}
 After evaluating the integrals and normalizing, the angular dipole transfer
 function is
 \begin{align}
   H_l^m(\mb{r}) \stackrel{(p)}{=} &\frac{N}{3}\left[2\text{jinc}_0^2(\nu_cr) + \left(\frac{\text{NA}}{n_o}\right)^2\text{jinc}_1^2(\nu_c r)\right]\Lambda_0\delta_{\ell0}\delta_{m0} + \nonumber\\& \frac{N}{3}\left[-2\text{jinc}_0^2(\nu_c r) + 2\left(\frac{\text{NA}}{n_o}\right)^2\text{jinc}_1^2(\nu_c r)\right]\Lambda_2\delta_{\ell2}\delta_{m0},
 \end{align}
where $\Lambda_{\ell} = \sqrt{4\pi/(2\ell + 1)}$. 

\begin{figure}[ht]
 \centering
   \centering
   \includegraphics[scale=0.8]{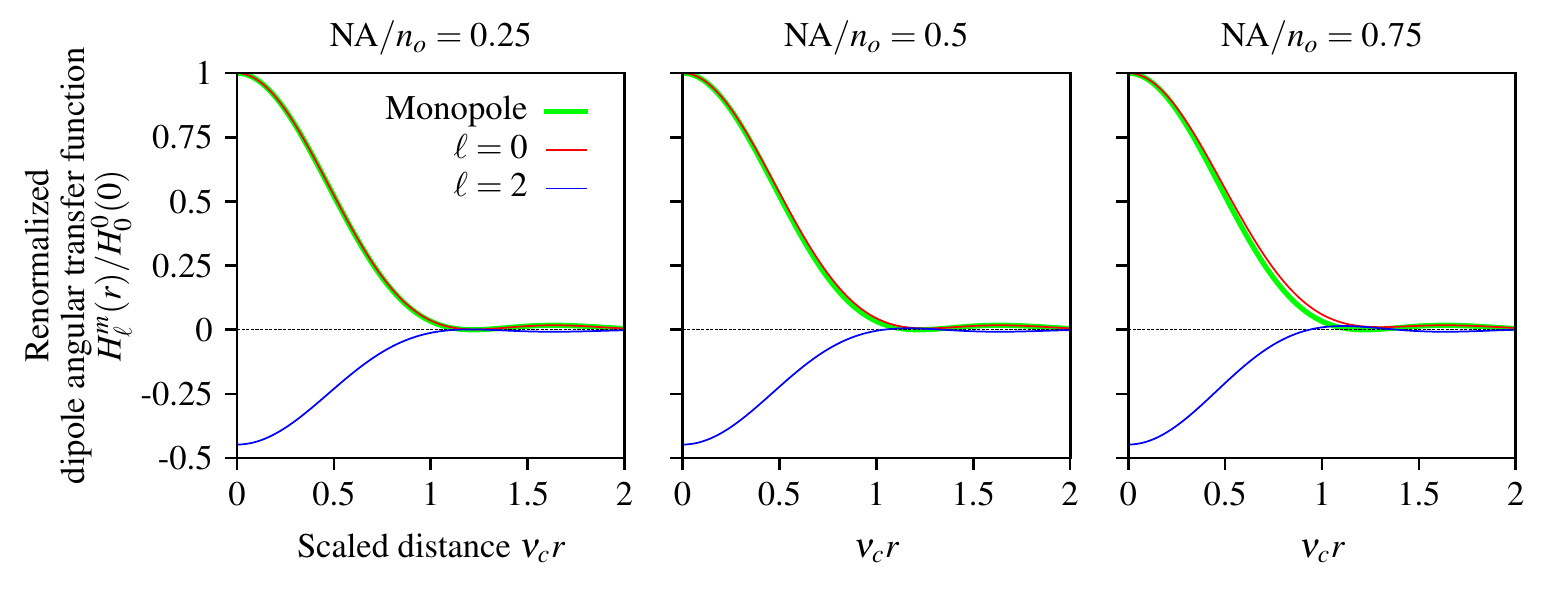}
   \caption{Paraxial dipole angular transfer function in terms of a scaled
     radial detection coordinate $\nu_c r$, the spherical harmonic degree
     $\ell$, and $\text{NA}/n_o$. Angularly uniform distributions of dipoles
     $\ell=0$ generate a spatial pattern that is similar but not identical to
     the Airy disk created by a monopole (\textcolor{green}{\textbf{green}}),
     and this discrepancy increases with the numerical aperture. $\ell=2$
     distributions have a negative response because $Y_2^0(\mh{s})$ is negative
     for transverse directions. As the numerical aperture increases, the
     relative contribution of positive axial dipoles in the $\ell=2$
     distribution increases.}
   \label{fig:atf}
 \end{figure}

In Fig. \ref{fig:atf} we plot the dipole angular transfer function for both
spherical harmonic terms and several numerical apertures. Note that the dipole
angular transfer function can be negative because the spherical harmonics can
take negative values. The $\ell=0$ term shows that angularly uniform
distributions of dipoles create spatial irradiance patterns that are similar but
not identical to the Airy disk, while the $\ell=2$ term shows a negative pattern
because of the large contribution of the transverse negative values in the
$Y_2^0$ spherical harmonic.

 To demonstrate the dipole angular transfer function we simulate a set of
 equally spaced fluorophore distributions with varying orientation and angular
 distributions
 \begin{align}
   f_{(ph3)}(r_x, r_y, \vartheta) = \sum_{j=0}^3 \sum_{k=0}^3 \delta\left(r_x - j\right)\, \delta\left(r_y - k\right)\, f_{\text{(cone)}}\left(\vartheta, \varphi; \vartheta_j, 0, \Delta_k\right),\label{eq:phantom3}
 \end{align}
 where
 \begin{align}
   f_{\text{(cone)}}(\so; \so', \Delta) = f_{\text{(cone)}}(\vartheta, \varphi; \vartheta', \varphi', \Delta) = \frac{1}{4\pi(1 - \cos\Delta)}\Pi\left(\frac{\mh{s}\cdot\mh{s}'}{2\cos\Delta}\right)
 \end{align}
 is an angular double cone distribution with central direction $\mh{s}'$ and
 cone half-angle $\Delta$; $\vartheta_j = j\frac{\pi}{6}$; and
 $\Delta_k = k\frac{\pi}{6}$. Notice that when $\Delta = 0$ the angular double
 cone reduces to a single direction, and when $\Delta = \pi/2$ the angular
 double cone reduces to an angularly uniform distribution. Also notice that the
 double cone can model angular diffusion or the angular distribution of many
 fluorophores within a resolvable volume.
 
 Our first step towards the irradiance pattern is to calculate the dipole
 angular spectrum of the phantom. In Appendix \ref{sec:cone} we calculate the
 spherical Fourier transform of the double cone distribution
 $F^m_{\ell,\text{(cone)}}(\vartheta', \varphi'; \Delta)$ which we can use to
 express the dipole angular spectrum as
 \begin{align}
   F^{m}_{\ell,(ph3)}(r_x, r_y, \vartheta) = \sum_{j=0}^3 \sum_{k=0}^3 \delta\left(r_x - j\right)\, \delta\left(r_y - k\right)\, F^m_{\ell,\text{(cone)}}\left(\vartheta_j, 0, \Delta_k\right).\label{eq:phantom3spect}
 \end{align}

 To calculate the scaled irradiance we multiply the dipole angular spectrum
   by the dipole angular transfer function and sum over the dipoles and
   spherical harmonics
 \begin{align}
   g_{(ph3)}(r_x, r_y) = \sum_{\ell m}\sum_{j=0}^3 \sum_{k=0}^3 H_\ell^m\left(\sqrt{\left(r_x - j\right)^2 + \left(r_y - k\right)^2}\right) F^m_{\ell,\text{(cone)}}\left(\vartheta_j, 0, \Delta_k\right).\label{eq:phantom3irr}
 \end{align}

\begin{figure}[ht]
 \centering
   \centering
   \includegraphics[scale=0.8]{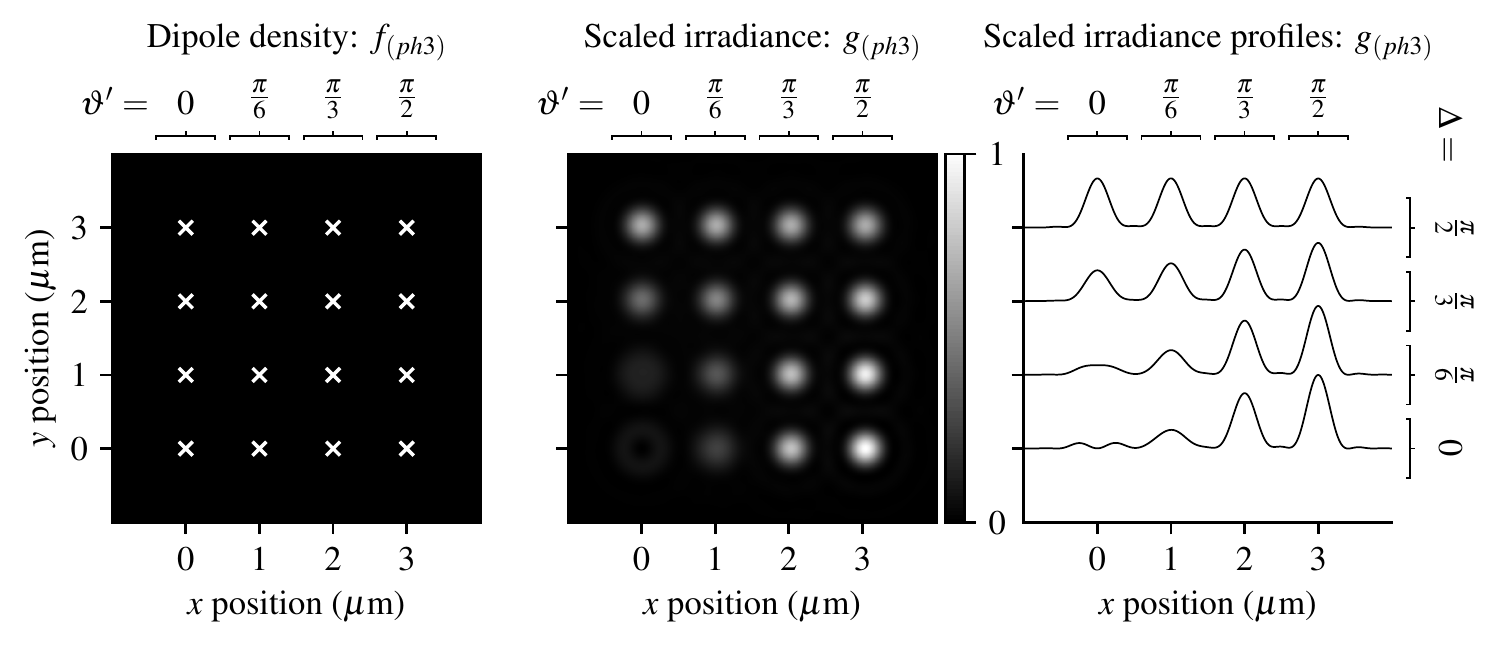}
   \caption{\textbf{Left:} A spatially sparse and angularly dense
     phantom---uniformly spaced double cone distributions of fluorophores with
     varying central direction (increasing $\vartheta'$ from left to right) and
     varying cone half-angle (increasing $\Delta$ from bottom to top).
     \textbf{Center:} Scaled irradiance for an imaging system with
     $\text{NA} = 0.75$, $\lambda = 500\,\text{nm}$, and $n_o = 1.33$ sampled at
     $20\times$ the Nyquist rate. \textbf{Right:} $x$ profiles through the
     scaled irradiance. Small cone angles have irradiance patterns that vary
     with the central direction, while larger cones angles have increasingly
     uniform irradiance patterns that hide angular information.}
   \label{fig:ph3}
 \end{figure}

 In Fig. \ref{fig:ph3} we plot the phantom and scaled irradiance with the same
 imaging parameters as the previous sections. For small cone angles the
 irradiance patterns are similar to the point sources in the previous sections,
 while larger cone angles create increasingly uniform irradiance patterns that
 hide the angular information about the distributions.

 \subsubsection{Paraxial dipole spatio-angular transfer function}
 We can calculate the dipole spatio-angular transfer function by taking the
 spatial Fourier transform of the dipole angular transfer function (or the
 spherical Fourier transform of the dipole spatial transfer function) to find
 that
 \begin{align}
\msf{H}_{\ell}^m(\bv) \stackrel{(p)}{=} &\frac{N}{3\nu_c^2}\left[2\text{chat}_0\left(\frac{\nu}{\nu_c}\right) + \left(\frac{\text{NA}}{n_o}\right)^2\text{chat}_1\left(\frac{\nu}{\nu_c}\right)\right]\Lambda_0\delta_{\ell0}\delta_{m0}+\nonumber\\ &\frac{N}{3\nu_c^2}\left[-2\text{chat}_0\left(\frac{\nu}{\nu_c}\right) + 2\left(\frac{\text{NA}}{n_o}\right)^2\text{chat}_1\left(\frac{\nu}{\nu_c}\right)\right]\Lambda_2\delta_{\ell2}\delta_{m0}.
 \end{align}

\begin{figure}[ht]
 \centering
   \centering
   \includegraphics[scale=0.8]{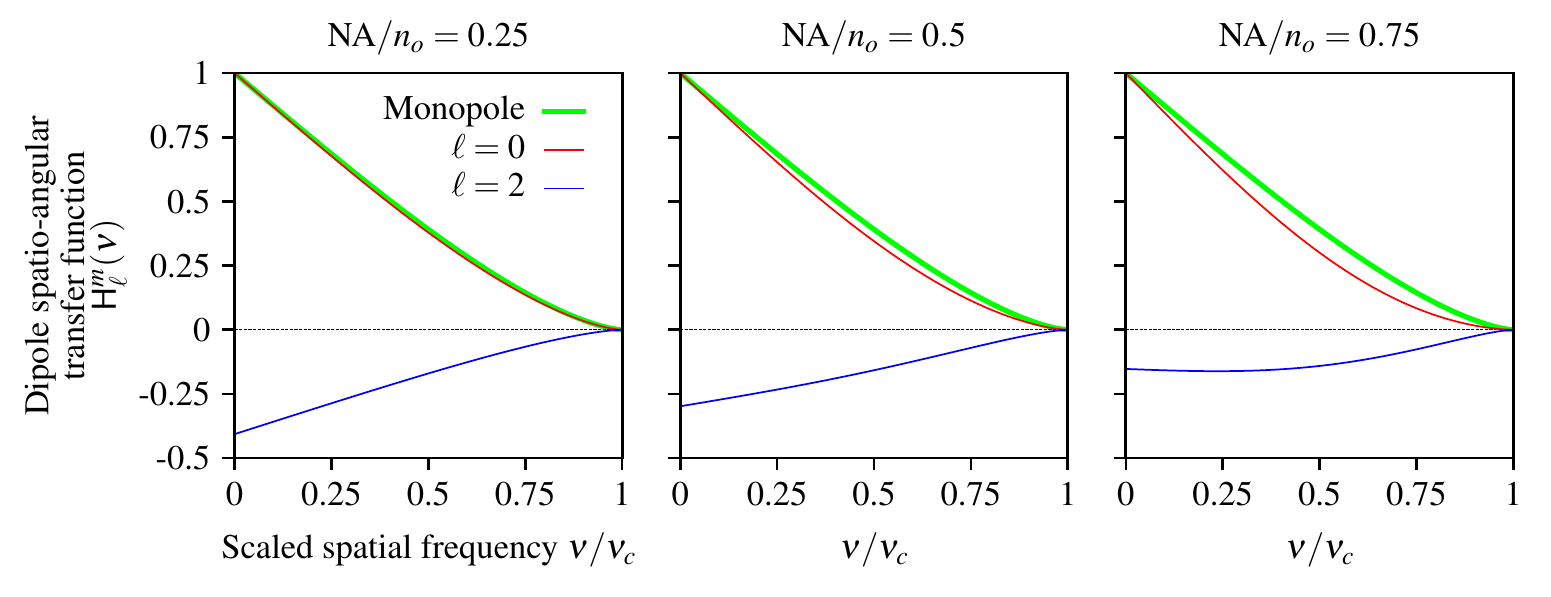}
   \caption{Spatio-angular dipole transfer function as a function of the scaled
     spatial frequency $\nu/\nu_c$, the spherical harmonic degree $\ell$, and
     $\text{NA}/n_o$. When the numerical aperture is small the transverse
     dipoles contribute the most to the signal which gives rise to a positive
     $\ell=0$ component and a negative $\ell=2$ component. As the numerical
     aperture increases, the relative contribution of axial dipoles increases
     and the $\ell=2$ component becomes less negative. Additionally, we plot the
     monopole transfer function (\textcolor{green}{\textbf{green}}) and observe
     that the $\ell=0$ term is similar but not identical to the monopole
     transfer function, and this discrepancy increases with the numerical aperture.}
   \label{fig:dsatf}
 \end{figure}

 In Fig. \ref{fig:dsatf} we plot the dipole spatio-angular transfer function for
 both spherical harmonic terms and several numerical apertures. The $\ell=0$
 term shows that an angularly uniform distribution of dipoles has a transfer
 function that is similar but not identical to the monopole transfer function
 with high frequencies increasingly suppressed as the numerical aperture
 increases. The $\ell=2$ term shows a negative pattern because of the large
 contribution of the transverse negative values in the $Y_2^0$ spherical
 harmonic. As the numerical aperture increases the relative contribution of the
 positive axial values increases and the $\ell=2$ term becomes less negative.
 
 To demonstrate the spatio-angular transfer function, we simulate a set of
 equally spaced disks of fluorophores with varying radius and angular
 distributions
 \begin{align}
   f_{(ph4)}(r_x, r_y, \vartheta, \varphi) = \sum_{j=0}^3 \sum_{k=0}^3 \frac{1}{D_k^2}\Pi\left(\frac{1}{D_k}\sqrt{\left(r_x - j\right)^2 + \left(r_y - k\right)^2}\right)f_{\text{(cone)}}\left(\vartheta, \varphi; \frac{\pi}{2}, 0, \Delta_j\right),\label{eq:phantom4}
 \end{align}
 where $D_k = 0.15(1+k)\, \mu$m, and $\Delta_j = j\frac{\pi}{6}$. 
 
 Our first step towards calculating the irradiance pattern is to calculate the
 dipole spatio-angular spectrum given by the spatial Fourier transform of the dipole angular spectrum  
  \begin{align}
   \mathsf{F}_{\ell,(ph4)}^m(\nu_x, \nu_y) = \mc{F}_{\mbb{R}^2}\left\{\sum_{j=0}^3 \sum_{k=0}^3 \frac{1}{D_k^2}\Pi\left(\frac{1}{D_k}\sqrt{\left(r_x - j\right)^2 + \left(r_y - k\right)^2}\right)F_{\ell,\text{cone}}^m\left(\frac{\pi}{2}, 0, \Delta_j\right)\right\}.
  \end{align}
  
  To calculate the scaled irradiance we multiply the dipole spatio-angular spectrum by the dipole spatio-angular transfer function, sum over the spherical harmonics, then take an inverse Fourier transform
  \begin{align}
   g_{(ph4)}(r_x, r_y) = \mc{F}^{-1}_{\mbb{R}^2}\left\{\sum_{\ell m}\mathsf{H}_{\ell}^m(\nu_x, \nu_y)\mathsf{F}_{\ell,(ph4)}^m(\nu_x, \nu_y)\right\}.
  \end{align}

\begin{figure}[ht]
 \centering
   \centering
   \includegraphics[scale=0.8]{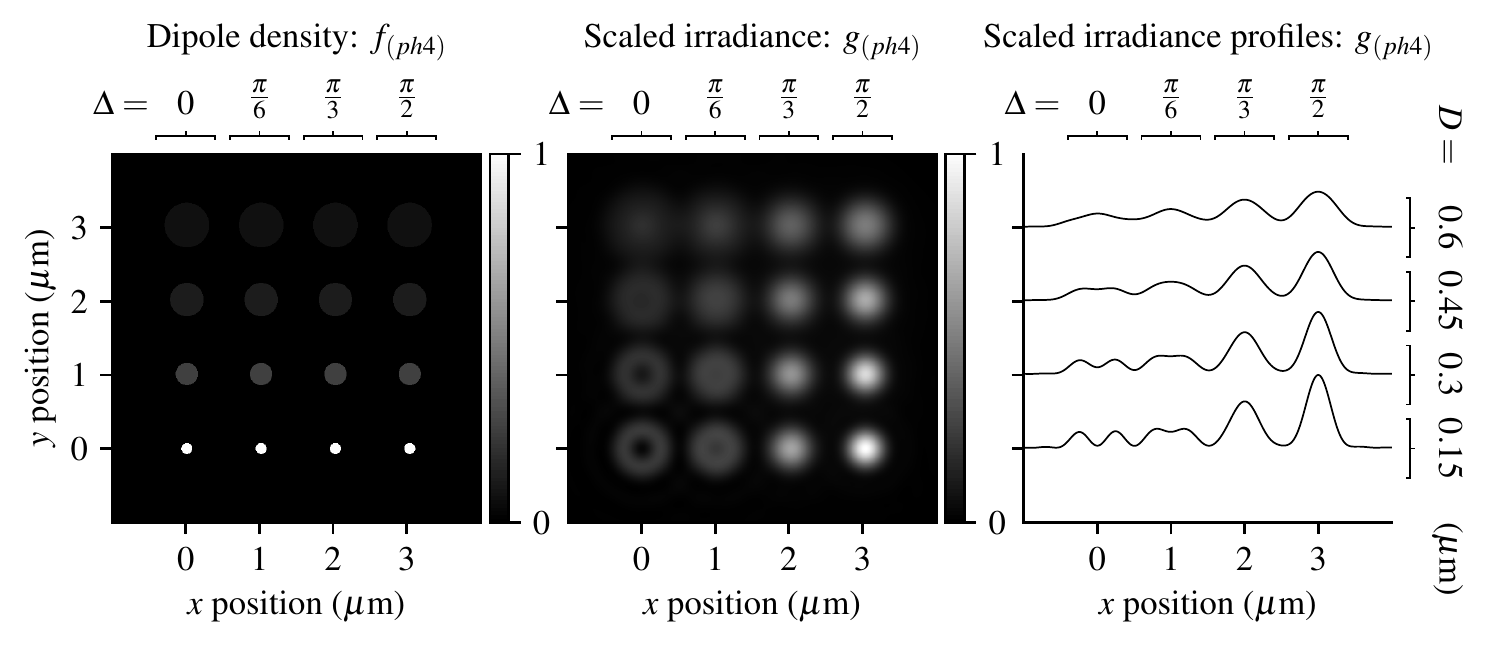}
   \caption{\textbf{Left:} A spatially and angularly dense phantom---uniformly
     spaced disks with varying size (increasing $D$ from bottom to top) and
     double cone half angle (increasing $\Delta$ from left to right)
     \textbf{Center:} Scaled irradiance for an imaging system with
     $\text{NA} = 0.75$, $\lambda = 500\,\text{nm}$, and $n_o = 1.33$ sampled at
     $20\times$ the Nyquist rate. \textbf{Right:} $x$ profiles through the
     scaled irradiance.}
   \label{fig:ph4}
 \end{figure}

  In Fig. \ref{fig:ph4} we plot the phantom and scaled irradiance with the
  same imaging parameters as the previous sections. Small cone angles and small
  disks create relatively unique irradiance patterns, while increasing the cone
  angle or disk size creates increasingly similar irradiance patterns. 
  
 \section{Discussion}\label{sec:discussion}
 \subsection{Comparing monopole and dipole models}
 The only case when the dipole and monopole transfer functions match exactly is
 when the sample consists of dipoles that are completely constrained to the
 transverse plane of a paraxial imaging system. Applying the monopole
 approximation in any other situation can lead to biased estimates of the
 fluorophore concentrations. To see how these biases manifest, consider the
   irradiance pattern created by an ensemble of dipoles oriented along the optic
   axis---see Figs. \ref{fig:ph1}, \ref{fig:ph2}, \ref{fig:ph3}, or \ref{fig:ph4}.
   Any reconstruction scheme that uses the monopole approximation would
   attribute the irradiance doughnut to a doughnut of monopoles instead of
   axially oriented dipoles, which is a clear example of a biased estimate caused
   by model mismatch.

   However, the common justifications for the monopole approximation---that the
   fluorophores are rotationally unconstrained or that
   there are many randomly
   oriented fluorophores in a resolvable volume---are good justifications in all
   but the highest SNR regimes. The effects of the dipole model become apparent
   in lower SNR regimes as the rotational constraints on the dipoles increase
   (assuming there are out-of-plane dipole components).
 
\subsection{What determines the angular bandwidth?}
Spatial imaging systems have a spatial bandwidth that characterizes the highest
spatial frequency that the system can transfer between object and data space.
Similarly, angular imaging systems have an angular bandwidth that characterizes
the highest angular frequency that the system can transfer, but in the angular case
there are two different types of angular bandwidths that we call the $\ell$- and
$m$-bandwidth. The $\ell$-bandwidth can be interpreted in a similar way to the
spatial bandwidth---it characterizes the smallest angular features that the
imaging system can measure. The $m$-bandwidth does not have a direct analog in
the spatial domain---it characterizes the angular uniformity of the imaging
system. If the $\ell$- and $m$-bandwidths are equal then the imaging system can
be said to have an $\textit{isotropic angular bandwidth}$.

The spatial bandwidth of a fluorescence microscope is well known to be
$\nu_c = \frac{2\text{NA}}{\lambda}$. In other words, we can increase the
spatial resolution of a fluorescence microscope by increasing the NA of the
instrument or by choosing a fluorophore with a shorter emission wavelength.
Similarly, the angular bandwidth of a fluorescence microscope depends on both
the instrument and the choice of fluorophore.

The microscope we considered in this work has an $\ell$-bandwidth of
$\ell_c = 2$ and an $m$-bandwidth of $m_c = 0$, so it does not have an isotropic
angular bandwidth. In future work we will consider several approaches to
improving the angular bandwidths in detail, but we briefly mention that
non-paraxial microscopes, microscopes with polarizers in the illumination or
detection paths, and multiview microscopes all have higher angular bandwidths
than the microscope considered here.

The angular bandwidth is also fluorophore dependent. Monopoles emit light
isotropically so they have an $\ell$-bandwidth of $\ell_c = 0$, while dipoles
have an $\ell$-bandwidth of $\ell_c = 2$, and higher-order excitation and
detection moments will have even higher bandwidths. Multi-photon excitation and
other non-linear methods can also increase the $\ell$-bandwidth
\cite{brasselet2011}.

\subsection{Towards more realistic models}
The theoretical model we presented in this work is an extreme simplification of
a real microscope. We have ignored the effects of thick samples,
refractive-index mismatch, aberration, scattering, finite fluorescence
lifetimes, and interactions between fluorophores among others. Because of this
long list of unconsidered effects, real experiments will likely require
extensions of the models developed here.

The dipole pupil function provides the simplest way to create more realistic
models from the simple model in this paper. Phase aberrations can be added to
the dipole pupil function with Zernike polynomials, and refractive index
boundaries can be modeled by applying the work of Gibson and Lanni to the dipole
pupil function \cite{gibson89}. These additions will model phase aberrations,
but modeling polarization aberrations will also be necessary, and we anticipate
that vector Zernike polynomials and the Jones pupil \cite{zhao2007, xu2015,
  chipman1989} will be essential tools for modeling dipole imaging systems. We
plan to use the dipole pupil function to include the effects of non-paraxial
objectives, polarizers, and defocus in future papers of this series.

The dipole pupil function also provides an enormous set of design opportunities.
The dipole imaging problem may benefit from spatially varying diattenuating and
birefrigent masks---a much larger set of possibilities than the well-explored
design space of amplitude and phase masks. The dipole pupil function is a step
towards Green's tensor engineering \cite{agrawal2012}, and the dipole transfer
functions provide a strong framework for evaluating dipole imaging designs.

In the simple case considered here we focused on the emission path of the
microscope, but the excitation path is equally important. Complete models will
need to consider the spatio-angular dependence of excitation. Zhenghao et. al.
\cite{zhanghao2017} have taken steps in this direction by considering polarized
structured illumination microscopy. Rotational dynamics and the fluorescence
lifetime are also important to consider when incorporating models of the
excitation process \cite{lew2013, zhang2018, zhang2018-2}.

\section{Conclusions}

We have calculated the monopole and dipole transfer functions for paraxial $4f$
imaging systems and demonstrated these transfer functions with efficient
simulations. We found that the monopole and scalar approximations are good
approximations when the sample consists of unconstrained rotating fluorophores
or many randomly oriented fluorophores within a resolvable volume. We also found
that dipole and vector optics effects become larger as rotational order
increases, and in these cases the dipole transfer functions become valuable
tools.

\section*{Funding}
National Institute of Health (NIH) (R01GM114274, R01EB017293).

\section*{Acknowledgments}
TC was supported by a University of Chicago Biological Sciences Division
Graduate Fellowship, and PL was supported by a Marine Biological Laboratory
Whitman Center Fellowship. Support for this work was provided by the Intramural
Research Programs of the National Institute of Biomedical Imaging and
Bioengineering.

\section*{Disclosures}
The authors declare that there are no conflicts of interest related to this article.

\appendix

\section{Relationships between special functions} \label{sec:special}
Our first task is to show that 
\begin{align}
  i^{n}\left\{\substack{
  \text{exp}(in\phi_r)\\
    \cos(n\phi_r)\\
    \sin(n\phi_r)
  }\right\}\text{jinc}_{n}(r) &\stackrel{\mathcal{F}_{\mbb{R}^2}}{\longrightarrow}   (2\nu)^{n}\left\{\substack{
    \text{exp}(in\phi_{\nu})\\
    \cos(n\phi_{\nu})\\
    \sin(n\phi_{\nu})
  }\right\}\Pi(\nu). \label{eq:jincrect3}
  \end{align}
  Writing the inverse Fourier transform in polar coordinates yields
  \begin{align}
    =2^n\int_0^{1/2}d\nu\, \nu^{n+1}\int_0^{2\pi}d\phi_\nu\left\{\substack{
    \text{exp}(in\phi_{\nu})\\
    \cos(n\phi_{\nu})\\
    \sin(n\phi_{\nu})
  }\right\}\text{exp}[2\pi i\nu r\cos(\phi_\nu - \phi_r)].
  \end{align}
  The azimuthal integral can be evaluated in terms of an $n^{th}$ order Bessel
  function (for the complex case see \cite[ch. 4.111]{barrett2004}).
  \begin{align}
    =2^n 2\pi i^n \left\{\substack{
    \text{exp}(in\phi_{r})\\
    \cos(n\phi_r)\\
    \sin(n\phi_r)
  }\right\}\int_0^{1/2}d\nu\, \nu^{n+1}J_n(2\pi\nu r).
  \end{align}
  We can use the following identity \cite[ch.~6.561-5]{gradshteyn2007}
  \begin{align}
    \int_0^1 du\, u^{n + 1}J_{n}(au) = a^{-1}J_{n + 1}(a)
  \end{align}
  with a change of variable $u = 2\nu$ to find the final result
  \begin{align}
    &=2^n 2\pi i^n \left\{\substack{
    \text{exp}(in\phi_{r})\\
    \cos(n\phi_r)\\
    \sin(n\phi_r)
  }\right\}\int_0^{1}\frac{du}{2}\, \left(\frac{u}{2}\right)^{n+1}J_n(\pi u r)
    =i^n \left\{\substack{
    \text{exp}(in\phi_{r})\\
    \cos(n\phi_r)\\
    \sin(n\phi_r)
  }\right\}\frac{J_{n+1}(\pi r)}{2r}
    =i^n \left\{\substack{
    \text{exp}(in\phi_{r})\\
    \cos(n\phi_r)\\
    \sin(n\phi_r)
  }\right\}\text{jinc}_{n}(r).
  \end{align}

\begin{figure}[ht]
 \centering
   \centering
   \includegraphics[width = 0.45\textwidth]{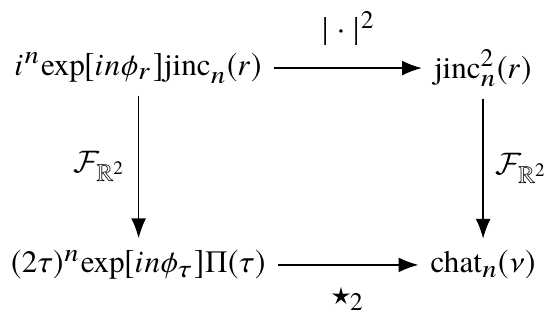}
   \caption{The relationships between special functions. The chat functions are
     defined as the two-dimensional Fourier transform of the squared jinc
     functions, and they can be calculated with the two-dimensional complex
     autocorrelations (denoted by $\star_2$) of the complex-weighted rectangle
     functions.}
   \label{fig:special}
 \end{figure}

\begin{figure}[ht]
 \centering
   \centering
   \includegraphics[width = 1.0\textwidth]{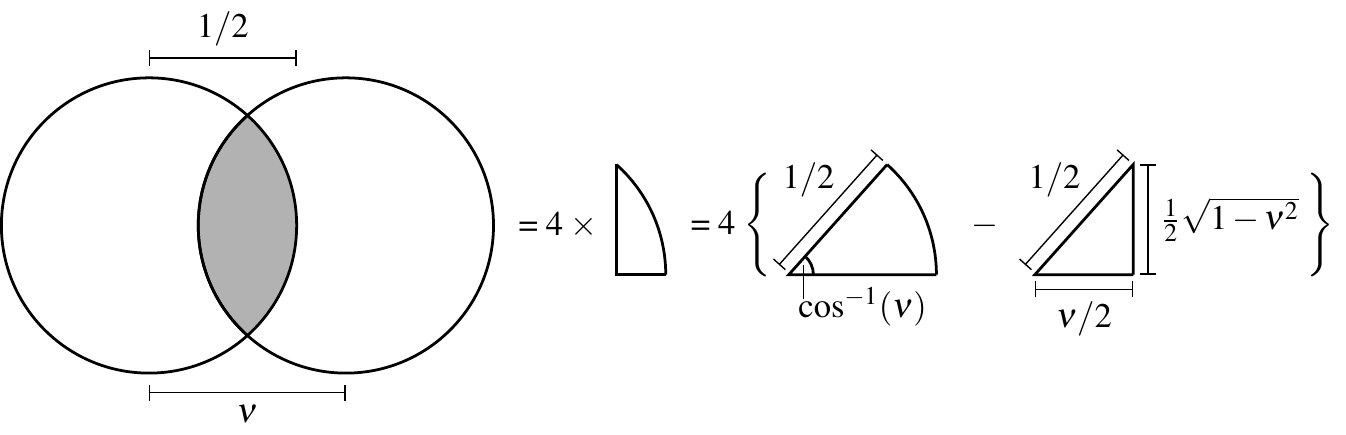}
   \caption{Geometric construction for evaluating the autocorrelation. We need
     to integrate over the overlapping region of two circles with radius $1/2$
     and distance $\nu$ between their centers. The region is four times the
     difference in area between a sector of angle $\cos^{-1}(\nu)$ and radius
     $1/2$ and a right triangle with base $\nu/2$ and hypotenuse $1/2$.}
   \label{fig:geometry}
 \end{figure}

We can use the relationship in Eq. \eqref{eq:jincrect3} to express the chat
functions in terms of a complex autocorrelation---see the diagram in Fig. \ref{fig:special}. Starting with the definition of
the $n^{th}$-order chat function
\begin{align}
  \text{chat}_{n}(\nu) = \int_{\mbb{R}^2}d\mb{r}\, \text{jinc}_{n}^2(|\mb{r}|)\, \text{exp}\left[-2\pi i\mb{r}\nu\right],  \label{eq:ftA}
\end{align}
we can rewrite the integrand in terms of the absolute square of a simpler
function with a known Fourier transform
\begin{align}
  \text{chat}_{n}(\nu) = \int_{\mbb{R}^2}d\mb{r}\, |t_n(\mb{r})|^2\, \text{exp}\left[-2\pi i\mb{r}\nu\right].  \label{eq:ftB}
\end{align}
\begin{align}
  t_n(\mb{r}) = i^n\text{exp}[in\phi_r]\text{jinc}_{n}(r).
\end{align}
Now we can apply the autocorrelation theorem to rewrite the Fourier transform as
\begin{align}
  \text{chat}_{n}(\nu) = \int_{\mbb{R}^2}d\bs{\tau}\, T_n(\bs{\tau})T^*_n(\bs{\tau} - \nu), \label{eq:autocorr}
\end{align}
where the function to be autocorrelated can be found with the help of Eq. \eqref{eq:jincrect3}
\begin{align}
  T_n(\bs{\tau}) = \int_{\mbb{R}^2}d\mb{r}\, t_n(\mb{r})\text{exp}\left[-2\pi i\mb{r}\cdot\bs{\tau}\right] = (2\tau)^n\text{exp}[in\phi_\tau]\Pi\left(\tau\right). \label{eq:tau}
\end{align}
It will be more convenient to set up the autocorrelation in Cartesian coordinates 
\begin{align}
  T_n(\bs{\tau}) &= 2^n(\tau_x + i\tau_y)^{n}\Pi\left(\sqrt{\tau_x^2 + \tau_y^2}\right). \label{eq:tn}
\end{align}
Plugging Eq. \eqref{eq:tn} into Eq. \eqref{eq:autocorr} gives
\begin{align}
  \text{chat}_{n}(\nu) = 4^n\int_{\mbb{R}^2}d\bs{\tau}\, (\tau_x^2 + \tau_y^2 - \nu\tau_x)^{n}\Pi\left(\sqrt{\tau_x^2 + \tau_y^2}\right)\Pi\left(\sqrt{(\tau_x - \nu)^2 + \tau_y^2}\right).
\end{align}
We can interpret the autocorrelation as an integral over a region of overlap
between a circle centered at the origin and a circle shifted to the right by
$\nu$ (a geometric lens). Using the construction in Fig. \ref{fig:geometry} we
can express this region as
\begin{align}
  \text{chat}_{n}(\nu) = 4^{n+1}\Bigg[&\int_0^{1/2}\tau d\tau\int_0^{\cos^{-1}\nu}d\phi_{\tau}(\tau^2 - \nu\tau\cos\phi_{\tau})^{n} -\nonumber \\ &\int_{0}^{\nu/2}d\tau_x\int_0^{\frac{\tau_x}{\nu}\sqrt{1 - \nu^2}}d\tau_y(\tau_x^2 + \tau_y^2 - \nu\tau_x)^{n}\Bigg]\Pi\left(\frac{\nu}{2}\right).
\end{align}

For $n=0$:
\begin{align}
  \text{chat}_0(\nu) &= 4\left[\int_0^{1/2}\tau d\tau\int_0^{\cos^{-1}\nu}d\phi_{\tau} - \int_{0}^{\nu/2}d\tau_x\int_0^{\frac{\tau_x}{\nu}\sqrt{1 - \nu^2}}d\tau_y\right]\Pi\left(\frac{\nu}{2}\right),\\
  \text{chat}_0(\nu) &= \frac{1}{2}\left[\cos^{-1}|\nu| - |\nu|\sqrt{1 - \nu^2}\right]\Pi\left(\frac{\nu}{2}\right),
\intertext{which is a well-known result \cite{goodman1996, mertz2009, bracewell2004}. For $n=1$:}
  \text{chat}_1(\nu) &= 16\Bigg[\int_0^{1/2}\tau d\tau\int_0^{\cos^{-1}\nu}d\phi_{\tau}(\tau^2 - \nu\tau\cos\phi_{\tau}) - \nonumber\\ &\qquad\quad\int_{0}^{\nu/2}d\tau_x\int_0^{\frac{\tau_x}{\nu}\sqrt{1 - \nu^2}}d\tau_y\, (\tau_x^2 + \tau_y^2 - \nu\tau_x)\Bigg]\Pi\left(\frac{\nu}{2}\right),\\
  \text{chat}_1(\nu) &= \frac{1}{2}\left[\cos^{-1}|\nu| - |\nu| (3 - 2\nu^2)\sqrt{1 - \nu^2}\right]\Pi\left(\frac{\nu}{2}\right). 
\end{align}

\section{Spherical Fourier transform of a double cone}\label{sec:cone}
In this appendix we evaluate the spherical Fourier transform of a
normalized double-cone angular distribution with central direction $\mh{s}'$ and
cone half-angle $\Delta$
\begin{align}
  f_{\text{(cone)}}(\mh{s}; \mh{s}', \Delta) = \frac{1}{4\pi(1 - \cos\Delta)}\Pi\left(\frac{\mh{s}\cdot\mh{s}'}{2\cos\Delta}\right). \label{eq:doublecone}
\end{align}
The spherical Fourier transform is
\begin{align}
  F_{\ell\text{(cone)}}^m(\mh{s}', \Delta) = \int_{\mbb{S}^2}d\mh{s}\, f_{\text{(cone)}}(\mh{s}; \mh{s}', \Delta)Y_\ell^{m*}(\mh{s}).
\end{align}

The limits of integration will be difficult to find unless we change coordinates
to exploit the axis of symmetry $\mh{s}'$. Since the spherical function is
rotationally symmetric about $\mh{s}'$ we can rotate the function so that the axis of
symmetry is aligned with $\mh{z}$ and multiply by $\sqrt{\frac{4\pi}{2l+1}}Y_\ell^{m*}(\mh{s}')$ to account for the rotation \cite{ramamoorthi2005}
\begin{align}
    F_{\ell\text{(cone)}}^m(\mh{s}', \Delta) = \sqrt{\frac{4\pi}{2l+1}}Y_\ell^{m*}(\mh{s}')\int_{\mbb{S}^2}d\mh{s}\, f_{\text{(cone)}}(\vartheta; \mh{z}, \Delta)Y_\ell^0(\mh{s}). 
\end{align}
In this coordinate system the double cone is independent of the azimuthal angle,
so we can evaluate the azimuthal integral and express the function in terms of an
integral over $\vartheta$:
\begin{align}
    F_{\ell\text{(cone)}}^m(\mh{s}', \Delta) = 2\pi Y_\ell^{m*}(\mh{s}')\int_{0}^\pi d\vartheta\, \sin\vartheta f_{\text{(cone)}}(\vartheta; \mh{z}, \Delta)P_\ell(\cos\vartheta). 
\end{align}
The function $f_{\text{(cone)}}(\vartheta; \mh{z}, \Delta)$ is only non-zero on
the intervals $\vartheta \in [0, \Delta]$ and
$\vartheta \in [\pi - \Delta, \pi]$ so
\begin{align}
    F_{\ell\text{(cone)}}^m(\mh{s}', \Delta) = \frac{Y_\ell^{m*}(\mh{s}')}{2(1 - \cos\Delta)}\left[\int_{0}^\Delta d\vartheta\, \sin\vartheta P_\ell(\cos\vartheta) + \int_{\pi - \Delta}^\pi d\vartheta\, \sin\vartheta P_\ell(\cos\vartheta)\right]. 
\end{align}
Applying a change of coordinates with $u = \cos\vartheta$ yields
\begin{align}
    F_{\ell\text{(cone)}}^m(\mh{s}', \Delta) = \frac{Y_\ell^{m*}(\mh{s}')}{2(1 - \cos\Delta)}\left[\int_{\cos\Delta}^1 d\vartheta\, P_\ell(u) + \int_{-1}^{-\cos\Delta} d\vartheta\, P_\ell(u)\right]. 
\end{align}
The Legendre polynomials $P_\ell(u)$ are even (odd) on the interval [-1, 1] when
$\ell$ is even (odd), so the pair of integrals will be identical when $\ell$ is
even and cancel when $\ell$ is odd. For even $\ell$,
\begin{align}
  F_{\ell\text{(cone)}}^m(\mh{s}', \Delta) = \frac{Y_\ell^{m*}(\mh{s}')}{1 - \cos\Delta}\int_{\cos\Delta}^1 d\vartheta\, P_\ell(u).
\end{align}
The integral evaluates to \cite[ch.~7.111]{gradshteyn2007}
\begin{align}
  \int_{\cos\Delta}^1 d\vartheta\, P_\ell(u) =
\begin{cases}
  1 - \cos\Delta\,, &\ell = 0\,,\\
  \sin\Delta\, P_l^{-1}(\cos\Delta)\,, &\text{else},
\end{cases}
\end{align}
where $P_l^{-1}(\cos\Delta)$ is the associated Legendre polynomial with order
$m=-1$, not an inverse Legendre polynomial. Bringing everything together
\begin{align}
  F_{\ell\text{(cone)}}^m(\mh{s}', \Delta) =
  \begin{cases}
    \sqrt{1/(4\pi)}, & \ell = 0,\\
    0, & \ell\, \text{odd},\\
    Y_\ell^{m*}(\mh{s}')\cot\left(\frac{\Delta}{2}\right)P_l^{-1}(\cos\Delta), & \ell > 0\,\, \text{even}.
  \end{cases}
\end{align}

\end{document}